\documentclass[prl,aps,showpacs,twocolumn,amsmath,amssymb]{revtex4}
\usepackage{graphicx}
\usepackage{amssymb,amsmath}

\usepackage{color}

\newcommand{\bino}[2]{\frac{#1!}{#2!(#1-#2)!}}
\newcommand{\Ucrit}{U^0_{\mathrm{crit}}}
\newcommand{\Umax}{U^0_{\mathrm{max}}}
\newcommand{\Vcrit}{\Delta V_{\mathrm{crit}}}
\newcommand{\Rb}{${}^{87}$Rb }
\newcommand{\nm}{\;\mathrm{nm}}
\newcommand{\kHz}{\;\mathrm{kHz}}

\begin{document}

\title{ Ultracold Bosons in a Tilted Multi-level Double-Well Potential}
\author{D.~R. Dounas-Frazer, A.~M. Hermundstad, and L.~D. Carr}
\affiliation{Department of Physics, Colorado School of Mines, 
Golden,
  CO, 80401}
\date{\today}

\begin{abstract}
The $N$-body problem in a double well requires new features for 
quantum information processing, macroscopic quantum superposition, 
and other fundamental studies of quantum many body physics in 
ultracold atoms.  One needs (a) tilt, and (b) to go beyond the  
single-particle ground state in each well, i.e., to two or more  
energy levels.  For (a), we show that a small potential difference 
between the wells, or tilt, causes the decoherence  of 
Schr\"odinger cat states.   However, these states reappear when 
the tilt can be compensated by atom-atom interactions; these tilted 
cat states constitute partial cats that are protected from 
decoherence by the many body wavefunction.  For (b), we provide 
explicit criteria for when two  energy levels are needed to 
describe the state space.   For typical experimental parameters, 
two levels are indeed required for creation of  cat states.
\end{abstract}


\maketitle

Recently, Bose-Einstein condensates (BECs) in double-well potentials 
have been the subject of diverse and exciting research.  For 
instance, such systems can be used to  search for deviations from Newtonian
gravity at small distances~\cite{Hall:2006, Schumm:2005} 
and to store and retrieve optical information~\cite{Ginsberg:2007}.  
Interest in these systems is not limited to practical applications; 
BECs in double-well potentials also provide an ideal medium for the 
study of fundamental quantum many-body phenomena, such as 
macroscopic quantum  tunneling~\cite{Shin:2005, Albiez:2005, 
Smerzi:1997, Milburn:1997, Zapata:1998, Menotti:2001} and
macroscopic superposition states, or ``Schr\"odinger cat'' 
states~\cite{Huang:2006, Mahmud:2005, Pitaevskii:2001, Louis:2001}.  
To describe these double-well systems, a Hubbard-like Hamiltonian 
has often been employed~\cite{Tonel:2005a, Spekkens:1998, 
Jaaskelainen:2005}.  However, these studies not only assume that the 
trapping potential is symmetric, but effects of excited levels are 
completely neglected.  {Variational methods have indicated that 
excited levels play 
a significant role~\cite{Menotti:2001, Alon:2007, Ananikian:2006}.}
Moreover, a {\it multi-level} picture of a  {\it
tilted} double-well potential is  \emph{required} for the creation 
of a quantum computer from neutral atoms~\cite{Sebby-Strabley:2006, 
Calarco:2004}, atom-chip-based gravity sensors~\cite{Hall:2006, 
Schumm:2005}, and  in the study of quantum transport 
phenomena~\cite{Niu:1996, Wilkinson:1996, Tomadin:2007}.
Thus, the current experimental context of double-well 
potentials has created an urgent need for a new theoretical 
analysis of the many-body double-well problem.

\begin{figure}\center
  \includegraphics[width=8.5cm]{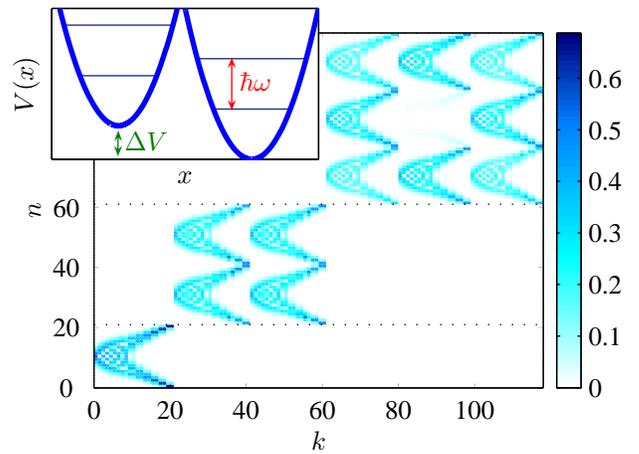}
  \caption{  (color online) {\it Energy eigenstates of a symmetric
      double well.}  
  Shown are the probability amplitudes $|c_n^{(k)}|$ for a small range of Fock 
  index $n$ and  energy eigenstate index $k$.  
  Amplitude is indicated by  hue. 
 The high barrier limit in which  macroscopic superposition states dominate
  the spectrum is depicted.  
 The total number of atoms is $N=20$; above
  the dotted line at Fock index $n=20$,
one atom occupies the higher energy level;
  above the dotted line at $n=60$ two atoms occupy the higher level; etc.
   Inset: schematic of the double-well potential. \label{fig:amp}}
\end{figure}

In this Letter we use a  two-level Hamiltonian to 
investigate the 
stationary states of a BEC in a tilted, one-dimensional (1D),   
double-well potential.  A schematic of our potential is shown in  
the inset of Fig.~\ref{fig:amp}.  In addition to  interaction with 
the electromagnetic vacuum~\cite{Huang:2006}, thermal 
effects~\cite{Pitaevskii:2001}, and dissipation~\cite{Louis:2001}, 
we find that tilt also causes the collapse of cat states when the 
barrier is high, an effect we term {\it potential decoherence}.  
Unlike for other forms of decoherence, such states reappear when the 
tilt  is compensated by atom-atom interactions.   We call this a 
{\it tunneling resonance} because tilt suppresses tunneling between 
wells except when these states 
reappear~\cite{Tonel:2005b,Dounas-Frazer:2006b}.  Finally, we present  novel, 
formal bounds for the use of a one-level approximation.  This 
approximation is typically thought to hold when the interaction 
energy is much smaller than the energy level 
difference~\cite{Ananikian:2006}.  However, we show that the effects 
of the excited energy level cannot be neglected even in this regime.

Past studies of the  $N$-body double-well problem have focused on 
symmetric traps with one allowed energy level.  Such systems map 
onto other physical problems, including a classical nonrigid 
pendulum~\cite{Smerzi:1997} and  the Lipkin-Meshkov-Glick model from 
nuclear physics~\cite{Vidal:2004}.  Moreover, a double-well 
potential coupled to a heat bath can be used to study  vibration 
modes in water molecules~\cite{Flach:2001}.   For the symmetric 
one-level model, tunneling dynamics~\cite{Smerzi:1997, Milburn:1997, 
Zapata:1998, Tonel:2005a} and the response to changes in 
experimental parameters such as barrier height have been  
analyzed~\cite{Jaaskelainen:2005, Huang:2006, Mahmud:2005, 
Tuchman:2006} in various regimes. The behavior of the ground state 
energy in a {\it tilted} double-well has been studied in the 
one-level approximation~\cite{Tonel:2005b}; however, applications 
such as quantum information processing~\cite{Sebby-Strabley:2006, 
Calarco:2004}  call for a two-level description.  In our study of 
the double-well potential, we relax two assumptions commonly made in 
previous studies of similar systems: the symmetric trap assumption 
and the one-level assumption.  This leads to two new energy scales 
in our problem.

An approximate two-level Hamiltonian for $N$ weakly interacting bosons 
in a tilted double-well potential is
\begin{equation}\label{eq:H}
  \hat{H} = \hat{H}^0 + \hat{H}^1 + \hat{H}^{01},
\end{equation}
where
\begin{eqnarray}
  \hat{H}^{\ell} &\equiv&{\textstyle-J^{\ell}\sum_{j\neq j'}\hat{b}_j^{\ell\dagger}
  \hat{b}_{j'}^{\ell}+U^{\ell}\sum_{j}\hat{n}_j^{\ell}\left(\hat{n}_j^{\ell}-1\right)}\nonumber\\
  &&+(\Delta V/2)\left(\hat{n}_L^{\ell}-\hat{n}_R^{\ell}\right)+E^{\ell}\left(\hat{n}_L^{\ell}+
  \hat{n}_R^{\ell}\right),
\end{eqnarray}
is the usual one-level Hamiltonian and
\begin{equation}\textstyle
\hat{H}^{01} \equiv 
U^{01}\sum_{j,\ell\neq\ell'}\left(2\hat{n}_j^{\ell}\hat{n}_j^{\ell'} 
+\hat{b}_j^{\ell\dagger}\hat{b}_{j}^{\ell\dagger}\hat{b}_j^{\ell'}\hat{b}_{j}^{\ell'}\right),\end{equation} 
couples the energy levels.  Here  
$[\hat{b}^{\ell}_j,\hat{b}^{\ell'\dagger}_{j'}]=\delta_{jj'} 
\delta_{\ell\ell'}$, 
$[\hat{b}^{\ell\dagger}_j,\hat{b}^{\ell'\dagger}_{j'}] 
=[\hat{b}^{\ell}_j,\hat{b}^{\ell'}_{j'}]=0$, and $\hat{n}^{\ell}_j 
\equiv \hat{b}^{\ell\dagger}_j\hat{b}^{\ell}_j$. 
 Equation~(\ref{eq:H}) can be derived from first principles quantum 
field theory for 
weakly interacting bosons at zero temperature~\cite{Rey:2004}.  The 
superscripts $\ell,\ell'\in\{0,1\}$ are the energy level indices, 
the subscripts $j,j'\in\{L,R\}$ are the well or site indices, 
$J^{\ell}$ are the tunneling energies, $U^{\ell}$ and $U^{01}$ 
are the interaction energies, $E^{\ell}$ is the energy of the
 $\ell$th excited level, and $\Delta V$ is the tilt. Setting 
$E^0=0$, the energy difference between levels, or level spacing, is 
$E^1\equiv\hbar\omega$. The extension of Eq.~(\ref{eq:H}) to an 
infinite number of sites leads to the two-band Bose-Hubbard 
Hamiltonian.

The two-level Hamiltonian allows for on-site interactions, tunneling 
between wells, and hopping between levels.  The tunneling terms 
$J^{\ell}$ allow single particles to tunnel between wells in the 
same level.  Pairs of particles in the same well interact with 
interaction energy $U^{\ell}$ if they are in the same energy level 
and $U^{01}$ if they are in different levels.  Interactions can be 
either repulsive, $U^{\ell}>0$, or attractive, $U^{\ell}<0$.  
{Furthermore, while single-atom transitions between energy levels are 
forbidden,} {\it 
two} atoms can hop together between energy levels with amplitude 
$U^{01}$.  Thus the energy levels are coupled by the inter-level 
interaction energy $U^{01}$.  Interactions between atoms in 
different wells are much smaller than $U^0$ and have been neglected 
in Eq.~(\ref{eq:H}). The parameters $J^{\ell}$, $\hbar\omega$, 
and $U^{\ell}$ are determined by overlap integrals of the localized
single-particle wavefunctions.  Left- and right-localized wavefunctions are 
constructed by superpositions of the appropriate symmetric and 
antisymmetric eigenfunctions of the single-particle Hamiltonian.  
{While the 
interaction and tunneling energies are independent parameters, 
$J^{\ell}$ and $\hbar\omega$ are not.} The 
tunneling and interaction energies satisfy $J^{0}\ll J^{1}\ll 
\hbar\omega$, { $U^{1}=(3/4)U^0$ and $U^{01}=(1/2)U^0$}.  In Figs.~1, 
2, and 3 we set  $J^0/\hbar\omega=4\times10^{-7}$ and 
$J^1/\hbar\omega=3\times10^{-5}$.  {Finally, we note that 
the use of single-particle wavefunctions and only a few levels
is appropriate to the
regime $|N U^0| \lesssim 2\hbar\omega$. When $|N U^0| \gtrsim 2\hbar\omega$ 
our approximation is inaccurate and alternative treatments become
necessary~\cite{Masiello:2006,Ananikian:2006,Alon:2007}.}

An arbitrary state vector in Fock space is given by
\begin{equation}
  |\Psi\rangle = \textstyle\sum_{n=0}^{\Omega-1}c_n|n\rangle, \:\:\:
  |n\rangle \in \{|n_L^0,\,n_R^0\rangle\otimes|n_L^1,\,n_R^1\rangle\},
\end{equation}
where $n_j^{\ell}$ is the number of particles in the 
 $\ell$th level of the 
$j$th well.  We require the total number of particles in the 
double-well, $N\equiv\sum_{j,\ell}n_j^{\ell}$, to be constant.  The 
Fock index $n$ increases with $n_L^0$, $n_L^0+n_L^1$, and $n_L^1$.   
 The dimension of the Hilbert space is $\Omega\equiv (N+3)(N+2)(N+1)/6$; in the one-level
approximation this reduces to simply $N+1$.  The energy eigenstates 
and eigenvalues of the Hamiltonian~(\ref{eq:H}) are given by 
 $\hat{H}|\phi_k\rangle = \varepsilon_k|\phi_k\rangle$,
for $0\leq k\leq\Omega-1$.  The  probability amplitudes are defined 
as $c_n^{(k)}\equiv\langle n|\phi_k\rangle$ for $0\leq 
n\leq\Omega-1$.  
{Throughout our treatment, we consider the regime $|NU^0|\ll2\hbar\omega$.
Furthermore, we work in the high barrier 
 limit $J^0\ll |U^0|$, as it is instrumental to the creation of stationary cat states.}    
 For simplicity, we restrict our discussion to repulsive interactions, $U^0>0$ and positive 
tilt, $\Delta V>0$.   However, our results hold for $U^0<0$ and 
$\Delta V<0$ as well.

\begin{figure}\center
  \includegraphics[width=8.5cm]{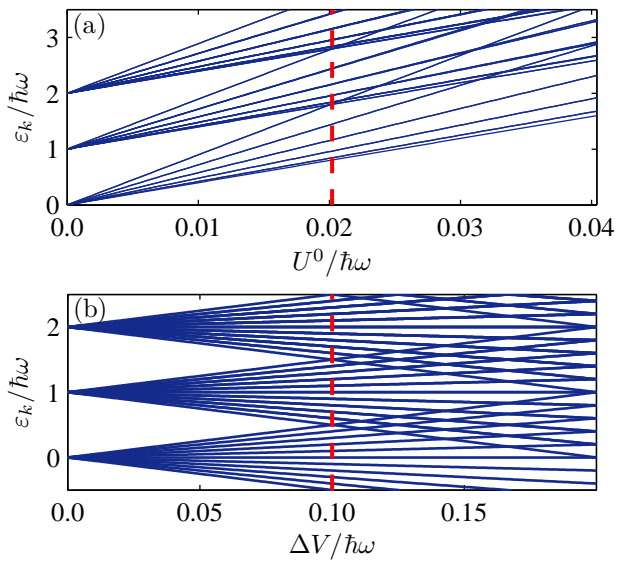}
  \caption{ {\it Breakdown of the one-level approximation.}  Energy eigenvalues versus (a) 
  interaction energy with $\Delta V = 0$ and (b) 
  tilt with $J^0/U^0 = 0.1$ for a ten-atom system.  The dashed red lines in 
  panels (a) and (b) correspond to $U^0=\Ucrit$ and $\Delta V = \Vcrit$, respectively.  
  Eigenvalue crossings cause states with nonzero occupation of the excited energy level 
  to emerge among the $N+1$ lowest-lying eigenstates. \label{fig:brk}}
\end{figure}

Characteristic eigenstate amplitudes are shown in Fig.~\ref{fig:amp} 
for a symmetric trap. For a given $n$, the number of particles in 
the excited level is equal to the number of horizontal dashed lines 
below $n$.  {In the case of Fig.~\ref{fig:amp}, mixing of levels is 
negligible.} The lowest excited group of 
eigenstates, for instance, corresponds to full occupation of the 
lowest level.  Likewise, the eigenvalues occur in $N+1$ groups 
approximately separated by the level spacing $\hbar\omega$.  As the 
number of particles,  the interaction energy, or the tilt increases, 
the spacing between groups  decreases,  as in Figs.~\ref{fig:brk}(a) and \ref{fig:brk}(b). 
In a symmetric trap, the first of many eigenvalue crossings occurs 
when the condition
\begin{equation}\label{eq:9}
  U^{0} \lesssim \Ucrit \equiv 2\hbar\omega/(N^2-1),
\end{equation}
is violated.  When $U^0 \gtrsim \Ucrit$, cat states with nonzero 
occupation of the excited level emerge among the first $N+1$ 
eigenstates and a two-level approximation must be used.  As the 
interactions increase, the lowest eigenstates include cat states 
with successively larger occupation of the excited level.  
Eventually, even the ground state will have significant 
contributions from the excited level.  {In a tilted potential, 
condition (\ref{eq:9}) becomes 
\begin{equation}\label{eq:99}
 \Delta V\!\lesssim\!\Vcrit\!\equiv\!\left\{\!\begin{array}{ll}
    \!\hbar\omega/N, \!&U^0\!\leq\!\Umax \\
    \!2U^0\!\!\left[\sqrt{1\!+\!2(\hbar\omega/U^0)}\!-\!N\right]\!\!,
    \!&U^0\!>\!\Umax
    \end{array}\right.\!,
\end{equation}
where we have defined $\Umax \equiv \Ucrit(N+1)/(4N)$.}   These results can be 
obtained formally in the limit of small 
tunneling.   In the opposite, 
noninteracting limit, we 
have developed a formal criterion involving the 
hopping as well~\cite{Dounas-Frazer:2007}:
$N < 1/2+(\hbar\omega - J^1)/(2J^0)$.

\begin{figure}\center
  \includegraphics[width=8.5cm]{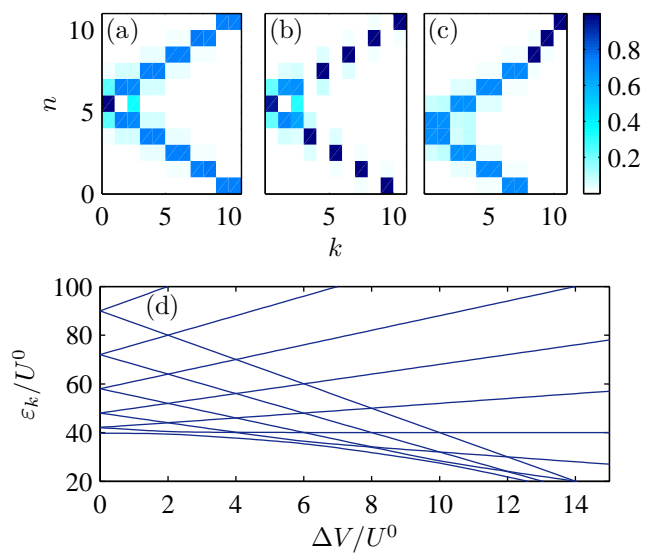}
  \caption{ {\it Potential decoherence and tunneling resonances.}  Eigenstate probability amplitudes for 
  a small range of Fock index $n$ and eigenstate index $k$ for $N=10$, $J^0/U^0 = 0.1$, and 
  (a) $\Delta V/U^0 = 0$, (b) $\Delta V/U^0 = 10^{-2}$, and (c) $\Delta V/U^0 = 6$. 
  (b)~A small tilt collapses the stationary cat states.  (d)~Avoided crossings in the energy 
  eigenvalues indicate a reappearance of  partial cat states, as in~(c). \label{fig:res}}
\end{figure}

While direct product states are not true eigenstates of the 
two-level Hamiltonian for $U^0>0$, the $N+1$ lowest excited 
eigenstates have negligible contributions from the excited level 
when conditions (\ref{eq:9}) and (\ref{eq:99}) are met.  In a 
symmetric potential, these states are of the form
 \begin{equation}\label{eq:4}
  |\phi_{\pm};\nu\rangle = |\Psi_{\pm};\nu\rangle\otimes|0,\,0\rangle,
\end{equation}
to $(N-2\nu-1)$th order in $J^0/U^0$ for $0\leq \nu< N/2$.  Here
 \begin{equation}
  |\Psi_{\pm};\nu\rangle \equiv (|\nu,\,N-\nu\rangle\pm|N-\nu,\nu\rangle)/\sqrt{2},
\end{equation}
represent {\it partial} cat states.  To $(N-2\nu)$th order in 
$J^0/U^0$, the energy difference  $\Delta \varepsilon_{\nu}$ of 
antisymmetric (-) and symmetric (+) pairs of states is 
\begin{equation}\label{eq:enu}
  \Delta\varepsilon_{\nu} = \frac{4U^0[J^0/(2U^0)]^{N-2\nu}(N-\nu)!}
  {\nu![(N-2\nu-1)!]^2}\,,
\end{equation}
which is a very small number. In the presence of a high barrier, the 
eigenstates therefore occur in nearly degenerate pairs of entangled 
states \cite{Huang:2006, Mahmud:2005} when the potential is 
symmetric.   Here  $\nu=0$ represents the {\it extreme} cat state in 
which all atoms simultaneously occupy both wells. Characteristic 
probability amplitudes for the $N+1$ lowest eigenstates are shown in 
Fig. \ref{fig:res}(a).

Because the level splitting between antisymmetric and symmetric 
pairs is so small, small perturbations can mix these states 
\cite{Huang:2006} and produce a localized state of the form  
$|\nu,\,N-\nu\rangle\otimes|0,\,0\rangle$.  Indeed, cat states are 
highly sensitive to tilt $\Delta V$,  as illustrated in 
Fig.~\ref{fig:res}(b). The cat states of Eq.~(\ref{eq:4}) are 
destroyed when
 \begin{equation}\label{eq:decoherence}
  \Delta V \gtrsim 2\Delta \varepsilon_{\nu}/ \left(N-2\nu\right).
\end{equation}
Small imperfections in the external potential thus constitute a 
source of quantum  decoherence.  In addition to dissipation and 
measurement, potential decoherence therefore poses a further 
difficulty in the engineering of cat states in experiments. Because 
{condition (\ref{eq:decoherence}) is minimized}
when  $\nu=0$, extreme cat 
states are the most sensitive to imperfections in the double-well, 
making them an unlikely candidate for experiments.  Partial cat 
states,  $\nu>0$, on the other hand, are more robust with respect to 
potential decoherence~\cite{footnote}.

Despite their fragility, cat states reappear periodically for 
certain values of the tilt.  Such resonances occur when
\begin{equation}\label{eq:res}\begin{array}{lr}
  \Delta V = \Delta V_p \equiv 2pU^0, & p \in \{1,\,2,\,\hdots,\,N-1\}.
\end{array}\end{equation}
In this case, the potential difference can be exactly compensated by 
the interaction of $p$ atoms in the lower well.  To  
$(N-2\nu-p-1)$th order in $J^0/U^0$, the energy eigenstates are 
partial cat states of the form 
\begin{align}
  |\phi_{\pm};\nu,\,p\rangle \!&=\! |\Psi_{\pm};\nu,\,p\rangle\otimes|0,\,0\rangle,\label{eq:pcat}\\
  |\Psi_{\pm};\nu,\,p\rangle \!&\equiv\! (|\nu,\,
  N\!-\!\nu\rangle\pm|N\!-\!\nu\!-\!p,\,\nu\!+\!p\rangle)/\sqrt{2},
\end{align}
for  $0\leq \nu<(N-p)/2$.  The reappearance of cat states is shown 
in Fig.~\ref{fig:res}(c) for $\Delta V = \Delta V_3$, i.e., the 
third resonance.  Because these states also occur in nearly 
degenerate pairs, the tunneling resonances are easily identified by 
avoided crossings in the energy eigenvalues, such as those displayed 
in Fig. \ref{fig:res}(d).  At odd integer values of  $\Delta V/U^0$, 
the eigenstates are maximally localized with the largest energy 
splitting between pairs of states.  Near a resonance, the 
eigenstates  become localized when
 \begin{equation}\label{eq:width}
  |\Delta V-\Delta V_p|\gtrsim2\Delta \varepsilon^p_{\nu}/(N-2\nu-p),
\end{equation}
where  $\Delta \varepsilon^p_{\nu} \propto (J^0/U^0)^{N-2\nu-p}$ is 
the energy difference between the states  $|\phi_-;\nu,\,p\rangle$ 
and $|\phi_+;\nu,\,p\rangle$.  For the special case  $\nu=0$, we 
find
\begin{equation}\label{eq:e0p}
  \Delta \varepsilon_0^p = \frac{4U^0[J^0/(2U^0)]^{N - p}(N-p)}{(N - p - 1)!}
    \sqrt{\bino{N}{p}},
\end{equation}
to $(N-p)$th order in $J^0/U^0$.  
 The understanding of tunneling resonances is vital in systems 
in which tilt is applied deliberately.{}


{} To demonstrate the robustness of the resonant cat states, we 
consider $N=100$ \Rb atoms in a 1D analog of the double-well potential of 
Ref.~\cite{Sebby-Strabley:2006}:
\begin{equation}
  V(x) = -v_1\cos^2(2kx)-v_2\cos^4(kx-\pi/4-\theta),
\end{equation}
for $kx\in[-\pi/4,3\pi/4]$, $v_2/v_1<2$, and $\theta\in[0,\pi/4]$.  Here $k=2\pi/\lambda$ where $\lambda = 810\nm$. We set $v_1=v_2/0.15=106E_r$ where $E_r/\hbar=22\kHz$ is the recoil energy of the lattice~\cite{detail}.  The radial trapping frequency is $\omega_{\perp} = 3.2\kHz$.   For such a potential, $\hbar\omega/E_r = 36$, $NU^0/(2\hbar\omega) = 0.10$, and $J^0/U^0=0.10$. The critical interaction energy and tilt are $\Ucrit/U^0 = 0.10$ and $\Vcrit/E_r = 0.36$, respectively.  In a symmetric trap, $\theta = 0$, the extreme cat states $|\phi_{\pm};0\rangle$ will collapse for deviations in the tilt on the order of $10^{-287} E_r$.  However, at the 98th resonance, i.e., when $\Delta V = \Delta V_{98} = 14E_r$, the few-atom superposition state $|\phi_{\pm};0,\,98\rangle$ can withstand deviations in the tilt up to $\pm0.098 E_r$.   The 98th resonance corresponds to $\theta = 0.52\pm1.2\%$~\cite{detail2}.  Clearly, the many-body wavefunction protects partial cat states; a superposition state of a few atoms \emph{without} the presence of a many-body cushion is not experimentally realistic~\cite{Dounas-Frazer:2006b}.

In conclusion, we have shown that not one, but three energy scales 
are required to describe macroscopic superposition states of 
ultracold atoms in a double well: hopping, tilt,  and energy 
level spacing, all in ratio to interaction.  These energy scales 
arise naturally within the context of quantum information processing 
with neutral atoms, where they are \emph{required} for quantum 
operations. After solving the appropriate Hamiltonian, we showed 
that tilt is a prevalent source of quantum decoherence which cannot 
be neglected in experiments. Moreover, we provided simple, formally 
derived criteria  for when all three energy scales are required.  
Our treatment of the two-level double-well potential was restricted 
to 1D for simplicity. While all considerations made here also apply 
to 2D and 3D, there is one very important difference.  In the latter 
case, angular momentum associated with the excited state in each 
well introduces another quantum number, as we analyze in detail 
elsewhere~\cite{Dounas-Frazer:2007}.

We thank Charles Clark, Mark Edwards, Qian Niu, William 
Phillips, Trey Porto,  and William Reinhardt for useful discussions. 
 This work was supported by the National Science 
Foundation under Grant PHY-0547845.


\begin{thebibliography}{10}
\bibitem{detail2}
{Although  $U^0>\Ucrit$ and $\Delta V_{98}>\Vcrit$, tunneling resonances are still described by Eqs. (\ref{eq:res}) and (\ref{eq:width}).}
\end{thebibliography}

\end{document}